# Phase-reference intensity attack on continuous-variable quantum key distribution with a real local oscillator

Yun Shao,[1] Yang Li,[1] Heng Wang,[1] Yan Pan,[1] Yaodi Pi,[1] Yichen Zhang,[2] Wei Huang,[1, *] and Bingjie Xu[1,†]

[1]Science and Technology on Communication Security Laboratory, Institute of Southwestern Communication, Chengdu 610041, China

[2]State Key Laboratory of Information Photonics and Optical Communications, Beijing University of Posts and Telecommunications, Beijing 100876, China

*huangwei096505@aliyun.com

†xbjpku@pku.edu.cn

In practical continuous-variable quantum key distribution system using local local oscillator (LLO CV-QKD), the phase noise related to coherent detection and phase-reference pulse intensity that can be locally calibrated at the receiver side is considered to be trusted noise to improve the performance. However, if the intensity of the phase-reference pulse is not monitored precisely in real-time, a security loophole will be opened for the eavesdropper to launch attacks. To this end, we propose two practical phase-reference pulse intensity attack strategies, where Eve can reduce the trusted phase noise by manipulating the intensity of the phase-reference pulse, thereby hiding her attack on the signal pulse if the total excess noise remains unchanged. The performance of the LLO CV-QKD system under these attacks has been analyzed. We show that precisely monitoring the intensity of the phase-reference pulse in real-time is an essential countermeasure to prevent the proposed attacks. Moreover, an intensity-monitoring scheme for the phase-reference pulse is proposed to strengthen the security of the practical LLO CV-QKD system and make the trusted phase noise model more robust.

## I. INTRODUCTION

Continuous-variable quantum key distribution (CV-QKD) allows two remote communicating parties Alice and Bob to share a secret key encoded in continuous variables, such as quadratures of coherent or squeezed states [1], which has attracted extensive interest due to its low cost, high detection efficiency, and good compatibility with existing telecom components [2-4]. Theoretically, its unconditional security is guaranteed based on the laws of quantum mechanics, so that any eavesdropping could be discovered. In practice, however, owing to the deviations between the theoretical security proofs and the practical CV-QKD implementations, there existed some security loopholes which can be exploited by the eavesdropper Eve to mount attacks. These attacks include wavelength attack [5-7], local oscillator (LO) calibration attack [8], LO fluctuation attack [9], detector saturation attack [10], homodyne detector binding attack [11], polarization attack [12], laser damage attack [13], Trojan-horse attack [14, 15], etc.

The Gaussian-modulated coherent state (GMCS) protocol [16] is the most widely used CV-QKD protocol since coherent states are easy to prepare in practice. Recently, for the GMCS CV-QKD, a field test over a 50-km commercial fiber [17], and a long distance transmission over a 202.81-km ultra-low loss fiber have been demonstrated [18]. In the GMCS CV-QKD, a coherent state signal pulse is Gaussian-modulated on the sender side and multiplexed co-transmitted with a phase-reference pulse through the optical fiber, and demultiplexed on the receiver side. Subsequently, the quadrature variables of the signal pulse are measured by homodyne or heterodyne detections. Conventionally, in a transmitting LO CV-QKD scheme [19], in order to meet the requirements of coherent detection, a strong LO is transmitted together with the signal pulse through the insecure quantum channel. However, this implementation not only leaves rooms for Eve to launch attacks, but also results in the leakage of photons from the LO to the quantum signal [20].

Recently, an appealing local LO scheme for CV-QKD (LLO CV-QKD) [21, 22] has emerged as an interesting candidate because of its avoidance of transmitting LO through the insecure quantum channel, which can completely prevent the LO attacks and has a potential to realize ultra-high speed CV-QKD. So far, several LLO CV-QKD schemes have been investigated and experimentally demonstrated [23-26], with a secret key rate up to 7.04 Mbits/s [25] and 26.9 Mbits/s [26] in the asymptotic limit over 25-km and 15-km fiber, respectively. In the LLO CV-QKD protocol, the signal pulse is generated at Alice's side while the LO is generated using another independent laser on Bob's side. To establish a reliable phase reference between Alice's and Bob's independent lasers, a low intensity classical phase-reference pulse is sent along with the signal pulse from Alice to Bob. Both the signal pulse and the phase-reference pulse are generated from the same laser. At the receiver, Bob uses his own LO to perform heterodyne detection for the phase-reference to estimate the relative phase between the two free-running lasers. Nevertheless, a non-negligible phase noise is introduced during the phase compensation process, which leads to a reduction of the system performance. More recently, a trusted phase noise model was developed to improve the phase noise tolerance of the LLO CV-QKD scheme [27], in which the secret key rate and transmission distance of the system are significant improved in comparison to that under the conventional phase noise model.

In this paper, we show that the LLO CV-QKD scheme under the trusted phase noise model [27] without a high precision phase-reference intensity monitoring is also potentially vulnerable to a hacking attack due to the fact that the phase-reference pulse propagates on the insecure quantum channel. We propose two practical attack schemes by manipulating the intensity of the phase-reference pulse, namely, Eve can hide her Gaussian collective attack by increasing the phase-reference pulse intensity with a phase-insensitive amplifier, or Eve can attenuate the initial phase-reference intensity during the communication parties' calibration process and then increase it in the QKD process. We also quantitatively analyze the secret key rates of the LLO CV-QKD system under the phase-reference intensity fluctuations attacks. The simulation further shows that the security of the system can be severely compromised if the phase-reference intensity is not monitored in real-time.

This paper is organized as follows. In Section II, we review the trusted phase noise model for LLO CV-QKD protocol based on Gaussian modulated coherent states and heterodyne detection. In Section III, we propose two practical phase-reference pulse intensity attack schemes. In Section IV, we discuss the countermeasures against these attacks. Finally, the conclusion is given in Section V.

## II. THE TRUSTED PHASE NOISE MODEL

To review the trusted phase noise model for LLO CV-QKD protocol with phase-reference, without loss of generality, and for simplicity, we describe it in a time-polarization multiplexing system based on GMCS protocol and heterodyne detection [28]. A schematic illustration of the time-polarization multiplexing LLO CV-QKD scheme is shown in Fig. 1. Suppose that two identical heterodyne detectors are used to detect the signal pulse and the phase-reference pulse, respectively. The measurement results of the phase-reference pulse allow Bob to recover the phase rotation between the two free-running lasers. A phase compensation for the signal pulse can thus be performed to generate the secret key.

In a time-polarization multiplexing LLO CV-QKD system, the total excess noise can be expressed as [27, 28]

$$\xi_{\text{tot}} = \xi_0 + \xi_{\text{AM}} + \xi_{\text{LE}} + \xi_{\text{ADC}} + \xi_{\text{phase}} \qquad (1)$$

Here, $\xi_0$ is the system excess noise from unidentified or unprotected sources [19]. $\xi_{\text{AM}}$ is the modulation noise caused by modulator imperfection in preparing the coherent state signal pulse, which is modeled as [29] $\xi_{\text{AM}} = (E_{S\max}^A)^2 10^{-d_{dB}/10}$, with $E_{S\max}^A = |\alpha_{S\max}| \approx \sqrt{10V_A}$ [29] being the maximal amplitude of the signal pulse, and $V_A$ being the modulation variance. $\xi_{\text{LE}}$ is the leakage noise caused by the photon leakage from the phase-reference pulse to the signal pulse, giving [27, 28] $\xi_{\text{LE}} = 2(E_R^A)^2/(R_e + R_p)$, with $E_R^A = |\alpha_R^A|$ being the amplitude of the phase-reference pulse on Alice's side, $R_e$ and $R_p$ being the extinction ratios corresponding to the amplitude modulator and the PBS. $\xi_{\text{ADC}}$ is the quantization noise introduced by the analog-to-digital (ADC) converters imperfection, which satisfies the constraint [28] $\xi_{\text{ADC}} \geq (E_{S\max}^A)^2/(12 \times 2^n)$, with n being the quantization number of the ADC. The last term $\xi_{\text{phase}}$ is the phase noise that is due to the deviation of phase-rotation estimation value of the signal pulse from its actual value [29]. One can combine the typical procedure of the phase-rotation estimation to quantify the deviation.

More generally, the phase noise mainly consists of three parts, given by [27-29]

$$\xi_{\text{phase}} = \xi_{\text{drift}} + \xi_{\text{channel}} + \xi_{\text{error}} \quad (2)$$

The first term $\xi_{\text{drift}}$ stands for the relative phase drift noise, which is introduced by the phase drift of the signal pulse relative to the phase-reference pulse at emission. The second term $\xi_{\text{channel}}$ stands for the relative phase accumulated noise, which is caused by the relative phase accumulation between the signal pulse and the phase-reference pulse during their propagation on the channel. The third term $\xi_{\text{error}}$ stands for the phase-reference measurement noise, which is derived from the fundamental shot noise and channel loss as well as the detector imperfection on the heterodyne detection of the phase-reference pulse, given by [27-29]

$$\xi_{\text{error}} = V_A \left( \frac{\chi + 1}{E_R^2} \right) \quad (3)$$

where $E_R$ is the phase-reference pulse amplitude at Bob's side, and $\chi$ is the total noise added on the phase-reference pulse, which can be expressed as [27, 28]

$$\chi = \frac{1-T}{T} + \varepsilon_0 + \frac{2 - \eta + 2v_{el}}{T\eta} \quad (4)$$

Here $\varepsilon_0$ is excess noise in the quantum channel for the phase-reference pulse, $\eta$ and $v_{el}$ respectively represent the detection efficiency and electronic noise of Bob's detector.

The trusted detector noise model [19, 30-32], as an effective model, has been widely adopted in CV-QKD experiments. In this model, the detector is assumed to be trusted, that is, the detector can not be exploited by Eve to mount attacks. In the conventional phase noise model, all sources of the phase noise are considered untrusted. Therefore, for the quantum signal, the total channel added noise referred to the channel input can be modeled as $\chi_{\text{line}} = 1/T - 1 + \xi_{\text{tot}}$, the detection added noise referred to Bob's input is given by $\chi_{\text{het}} = (2 - \eta + 2v_{el})/\eta$, and the total added noise referred to the channel input can be written as $\chi_{\text{tot}} = \chi_{\text{line}} + \chi_{\text{het}}/T$.

Based on the trusted phase noise model [27] where the phase-reference measurement noise associated with the detector as well as the phase-reference intensity at Bob's side that can be locally calibrated is considered trusted, the total noise imposed on the phase-reference pulse can be decomposed as

$$\chi = \chi^U + \frac{\chi^T}{T} \quad (5)$$

where $\chi^U = 1/T - 1 + \varepsilon_0$ and $\chi^T = (2 - \eta + 2\nu_{el})/\eta$. The phase-reference measurement noise can thus be written as

$$\xi_{error} = \xi_{error}^U + \frac{\xi_{error}^T}{T} \quad (6)$$

where $\xi_{error}^U = V_A(1 + \chi^U)/E_R^2$ is considered untrusted, and $\xi_{error}^T = V_A\chi^T/E_R^2$ is considered trusted. For a time-polarization multiplexing LLO CV-QKD system, the phase noise is determined by the phase-reference measurement noise, i.e., $\xi_{phase} \approx \xi_{error}$ [27-29]. According to Eq. (1), the total excess noise can thus be expressed as

$$\xi_{tot} = \xi_0 + \xi_{AM} + \xi_{LE} + \xi_{ADC} + \xi_{error} \quad (7)$$

and the added noises in the trusted phase noise model satisfy [27]

$$\chi_{line}^T = \frac{1}{T} - 1 + \xi_0 + \xi_{AM} + \xi_{LE} + \xi_{ADC} + \xi_{error} - \frac{\xi_{error}^T}{T}, \quad (8)$$

$$\chi_{het}^T = \frac{2 - \eta + 2\nu_{el}}{\eta} + \xi_{error}^T, \quad (9)$$

$$\chi_{tot}^T = \chi_{line}^T + \frac{\chi_{het}^T}{T}. \quad (10)$$

As discussed in Ref. [27], the performance of the LLO CV-QKD system under the trusted phase noise model is increased considerably compared with that under the conventional phase noise model.

### III. PHASE-REFERENCE PULSE INTENSITY ATTACK

In the trusted phase noise model, it is assumed that the phase noise $\xi_{error}^T$ associated with the detector efficiency $\eta$ and the detector electronic noise $\nu_{el}$, as well as the real-time monitored phase-reference intensity $E_R^2$ at Bob's side that can be calibrated locally is considered trusted [27]. However, in a practical CV-QKD system, the intensity of weak phase-reference pulse fluctuates in time during the QKD process, which is difficult to be monitored precisely in real-time. Hence, it is no surprise that a potential threat is posed by the transmission of the phase-reference pulse through the unsecure quantum channel.

Provided that the total detection added noise is calibrated before the QKD run, and the phase-reference pulse intensity remains unchanged during the QKD process, according to Eqs. (3) and (9), if the intensity of the phase-reference pulse is manipulated by Eve during the QKD run while Bob still uses the previously measured power to estimate the trusted phase noise, the communication parties will obtain a false key rate. Based on the above security loophole, we propose two practically feasible phase-reference pulse intensity attack strategies. In the following, we present the attacks against the time-polarization multiplexing LLO CV-QKD system based on GMCS protocol and heterodyne detection under the trusted phase noise model.

### A. Intensity attack using a phase-insensitive amplifier

In Fig. 2, we depict the phase-reference pulse intensity attack strategy using a phase-insensitive amplifier. Notice that the conventional preparation-and-measure (PM) implementation of a Gaussian-modulated coherent state CV-QKD protocol is equivalent to the corresponding entanglement-based (EB) scheme [33-35]. However, the phase-reference pulse used in a practical scheme is not taken into account in the underlying EB scheme theoretical model. In a practical LLO CV-QKD system, Eve can attack the signal pulse as well as the phase-reference pulse as follows. First, she intercepts both the signal pulse and the phase-reference pulse at the channel input, and then separates the signal pulse from the phase-reference pulse and transmit them through her own two quantum channels, respectively. For the signal pulse, Eve can perform a general Gaussian collective attack [33-35] to gain information, meanwhile, an attack noise is introduced inevitably. For the phase-reference pulse (see the inserted plot inside the blue dashed box in Fig. 2), Eve introduces a phase-insensitive amplifier [36] to increase the intensity of the phase-reference pulse, thereby resulting in a decrease of the trusted part of the phase noise [see Eq. (6)]. One can find that Eve could hide her increased attacks noise on the signal pulse if the total excess noise $\xi_{\text{tot}}$ remains unchanged during the QKD process. Second, Eve recombines the signal pulse and the phase-reference pulse at the channel output. Obviously, Alice and Bob will overestimate the value of the trusted phase noise and underestimate the other excess noises components in the practical implementations. Consequently, the security of the practical LLO CV-QKD system is compromised.

As illustrated in Fig. 2, a practical phase-insensitive amplifier with an amplification factor $g$ is used by Eve to amplify the intensity of the phase-reference pulse. The non-degenerate optical parametric amplifier is modeled as the transformations $X_R \to \sqrt{g}X_R +$

$\sqrt{g-1}X_I$ [36], with S and I being the phase-reference pulse mode and an idler mode with variance $V_I = N$ [36], respectively. Therefore, a noise is introduced during the amplification of the phase-reference pulse. According to Ref. [36], when the amplifier is placed near the channel output, the amplification noise imposed on the phase-reference pulse referred to the channel input can be expressed as

$$\chi^A = \frac{(g-1)N}{gT}. \tag{11}$$

Thus, the total noise imposed on the phase-reference pulse can be written as [36]

$$\chi = \frac{1-T}{T} + \varepsilon_0 + \frac{(g-1)N}{gT} + \frac{2-\eta+2v_{el}}{T g \eta} = \chi^U + \chi^A + \frac{\chi^T}{gT}. \tag{12}$$

According to Eqs. (3)−(6) and (11), as the amplifier will inevitably introduce noise during its operation, the additional phase noise introduced by the amplifier referred to the channel input can be written as

$$\xi^A_{error} = \frac{V_A \chi^A}{E^2_{ref}} = V_A \frac{(g-1)N}{gTE^2_R}. \tag{13}$$

Besides, the trusted part of the phase noise can be reduced by amplifying the intensity of the phase-reference pulse. From Eqs. (5) and (12), the reduction of the trusted part of the phase noise referred to the channel input is expressed as

$$\Delta \xi^T_{error} = V_A \frac{\chi^T}{TE^2_{ref}} - V_A \frac{\chi^T}{gTE^2_{ref}} = \frac{\xi^T_{error}}{T}\left(1 - \frac{1}{g}\right). \tag{14}$$

It is clearly that the phase-insensitive amplifier not only reduces the trusted phase noise, but also introduces additional phase noise, which is random for both the communicating parties and Eve. Meanwhile, Eve's attack on the signal pulse also introduces attack excess noise. In order to keep the total excess noise on Bob's measurements unchanged, Eve cannot use the overall reduction of the trusted phase noise to compensate her increased attack on the signal pulse. In fact, the reduction of the trusted phase noise should be equal to the sum of the additional phase noise and the increased attack excess noise. As a result, Eve's increased attack excess noise on the signal pulse should be

$$\xi_{attack} = \Delta\xi^T_{error} - \xi^A_{error} = \frac{\xi^T_{error}}{T}\left(1 - \frac{1}{g}\right) - V_A \frac{(g-1)N}{gTE^2_R}. \tag{15}$$

Hence, for the LLO CV-QKD system under the attack, the total channel added noise for the signal pulse referred to the channel input can be written as

$$\chi_{\text{line}}^{\text{attack}} = \frac{1}{T} - 1 + \xi_0 + \xi_{\text{AM}} + \xi_{\text{ADC}} + \xi_{\text{LE}} + \xi_{\text{error}} - \frac{\xi_{\text{error}}^T}{T} + \xi_{\text{attack}}, \quad (16)$$

and the total detection added noise referred to Bob's input thus satisfies

$$\chi_{\text{het}}^{\text{attack}} = \frac{2 - \eta + 2\nu_{\text{el}}}{\eta} + \xi_{\text{error}}^T - T\xi_{\text{attack}}, \quad (17)$$

The total noise referred to the channel input could be expressed as

$$X_{\text{tot}}^{\text{attack}} = \chi_{\text{line}}^{\text{attack}} + \frac{\chi_{\text{het}}^{\text{attack}}}{T} = \chi_{\text{tot}}^T. \quad (18)$$

Besides, it is interesting to note that the total excess noise obtained from the parameter estimation procedure is unchanged under the attack, Eve thus can hide her increased attack behavior on the quantum signal during the QKD run by appropriately amplifying the intensity of the phase-reference pulse.

The asymptotic secret key rate of LLO CV-QKD against collective attack, in the case of reverse reconciliation, can be expressed as $K = \beta I_{\text{AB}} - \chi_{\text{BE}}$ [30]. Here, $I_{\text{AB}}$ is the Shannon mutual information between Alice and Bob, $\chi_{\text{BE}}$ is the maximum information available to Eve bounded by the Helevo quantity, and $\beta$ is the reconciliation efficiency. The conventional secret key rate calculations for the heterodyne protocol is presented in detail in [27, 31, 36]. In Fig. 3, we plot the simulation results of secret key rate for the LLO CV-QKD system under some typical parameters. Compared to the conventional phase noise model, the trusted phase noise model can considerably improve the secret key rate and the transmission distance of the system. One can also find that for the trusted phase noise model under the phase-reference pulse intensity attack, the performance of the CV-QKD system is visibly degraded. It means that Eve's manipulation of the phase-reference pulse intensity can fully compromise the security of the LLO CV-QKD system. Notice that Eve's intercepted information of the signal pulse is proportional to the amplification factor g of the amplifier, and she can obtain partial or total key rate information shared between Alice and Bob. Taking the 30-km transmission distance as an example, for the LLO CV-QKD system under the trusted phase noise model without Bob's monitoring the phase-reference pulse intensity, Eve can obtain approximately 19% and 33% of the secret keys when the amplification factors are *g*=2 and *g*=10. In addition, Eve can obtain all the keys when the transmission distance exceeds 50-km and 45-km, corresponding to the amplification factors are *g*=2 and *g*=10. In this intensity attack scheme, Eve can mount the attack flexibly because she can freely choose the amplification factor and which part of the phase-reference pulses to be amplified.

### B. Intensity attack using a variable optical attenuator

Based on the scheme described above, one can find that utilizing the phase-insensitive amplifier will unavoidably introduce amplification noise, which can discount the efficiency of the intensity attack. Next, we propose another phase-reference intensity attack strategy without introducing additional excess noise. As shown in Fig. 4, Eve uses a variable-optical-attenuator (VOA) to attenuate the initial intensity of the phase-reference pulse between Alice and Bob to their calibrated value, and then reduce the attenuation ratio when Alice and Bob start the QKD protocol, so that the intensity of each phase-reference pulse is higher than the initial calibrated value. It's clear that this attack scheme is equivalent to a noiseless amplifier attack scheme in that the VOA does not add noise in this process. Here the phase-reference pulse intensity increases with the decrease of the attenuation ratio, while the change of the trusted phase noise is opposite. We introduce a parameter *r* as the ratio between the real value of the phase-reference pulse intensity during the QKD run and the initial calibrated value. Hence, in order to keep the total excess noise unchanged, Eve's increased attack excess noise on the signal pulse is equal to the reduction of the trusted phase noise, which can be obtained as

$$\xi_{\text{attack}} = \Delta\xi_{\text{error}}^T = \xi_{\text{error}}^T \left(1 - \frac{1}{r}\right), \qquad (19)$$

Additionally, expressions for the total channel added noise and the total detection added noise, as well as the total noise referred to the channel input are the same as that expressed in Eqs. (16) − (18). Combining this phase-reference pulse attack scheme with the calculations of [27, 31, 36], one can determine the secret key rate of the LLO CV-QKD system.

Figure 5 presents the simulation secret key rate results from the conventional phase noise model, the trusted noise model, as well as the trusted noise model under the phase-reference intensity attack using a variable attenuator. It shows that, within the attack scheme when $r$ takes the value $1/T$ (equivalent to transmitting the phase-reference pulse through a lossless channel), the performance of the system under the trusted phase noise model is severely reduced. Similarly, taking the 30-km transmission distance as an example, for the LLO CV-QKD system under the trusted phase noise model without Bob's monitoring the phase-reference pulse intensity, Eve can obtain approximately 45% of the secret keys. Moreover, Eve is able to obtain all the keys when the transmission distance exceeds 40-km. Apparently, with increasing $r$, the insecure region will be expanded and the secret key rate of the trusted phase noise model under attack will approach that in the conventional phase noise model. Since there is no excess noise associated with the use of attenuator, this attack scheme is more powerful than the previous one.

## IV. COUNTERMEASURE

As analyzed above, the trusted phase noise model manifests its importance as substantially promoting the secret key rate and transmission distance of the LLO CV-QKD system. However, Eve's attacks against the LLO CV-QKD by manipulating the phase-reference pulse intensity can severely compromise the security and performance of the system. Consequently, in a practical LLO CV-QKD system under the trusted phase noise model, monitoring the intensity of the phase-reference pulse in real-time is of great importance to guarantee the security of the implementations. In this case, Bob can use the real-time measured phase-reference pulse intensity to calibrate the trusted phase noise, i.e., $\xi_{error}^{T}$, to get a real secure key rate.

However, the characteristics of weak intensity of the phase-reference pulse has made it challenging to monitor its intensity precisely in real-time. In an efforts to overcome this challenge, a phase-reference pulse intensity monitoring scheme is presented in Fig. 6, in which an asymmetric beam splitter is used to split a small part of the phase-reference pulse, then amplifying the intensity appropriately with an amplifier, and measuring the power with a high-precision power meter. We can adopt the upper bound monitoring value of the phase-reference pulse intensity fluctuation to calibrate the trusted part of the phase noise to eliminate the overestimation of the secret key rate. We can find that monitoring the intensity of the phase-reference pulse precisely in real-time can resist the phase-reference pulse intensity attacks, and make the trusted phase noise model more robust.

## V. CONCLUSION

In conclusion, we have proposed two powerful and practical phase-reference pulse intensity attack strategies for LLO CV-QKD systems under the trusted phase noise model, by which Eve can compromise the security of the systems in the absence of effective countermeasures. The first strategy is to use a phase-insensitive amplifier to increase the phase-reference pulse intensity during the QKD process to reduce the trusted part of the phase noise. On the premise of keeping the total excess noise of Bob's measurement constant, Eve can hide her attack on the quantum signal by offsetting her increased attack noise with the reduced phase noise. The second strategy is to use a variable optical attenuator to attenuate the phase-reference pulse intensity to the initial calibrated value and then reduce the attenuation ratio during the QKD run, which is equivalent to increase the phase-reference pulse intensity with a noiseless amplifier. The simulation results show that both the attack schemes can seriously threaten the security of the LLO CV-QKD system. To enhance the security of the practical LLO CV-QKD system under the trusted phase noise model, we have discussed a countermeasure based on monitoring the intensity of the phase-reference pulse at Bob's side in real-time. In this case, the potential intensity side-channel attacks against the LLO CV-QKD system can be resisted. The present study will help make the trusted phase noise model more robust and promote practical applications of LLO CV-QKD.

## ACKNOWLEDGEMENTS

We acknowledge financial support from the National Natural Science Foundation of China (Grants No. 61771439, No. U19A2076, No. 61901425, No. 62171418, and No. 62101516), the National Cryptography Development Fund (Grant No. MMJJ20170120), the Sichuan Science and Technology Program (Grants No. 2019JDJQ0060 and No. 2020YFG0289), the Sichuan Application and Basic Research Funds (Grant No. 2020YJ0482), the Innovation Special Zone Funds (Grant No. 18-163-00-TS-004-040-01), and the Technology Innovation and Development Foundation of China Cyber Security

(Grant No. JSCX2021JC001).

# Figures and Figure captions:

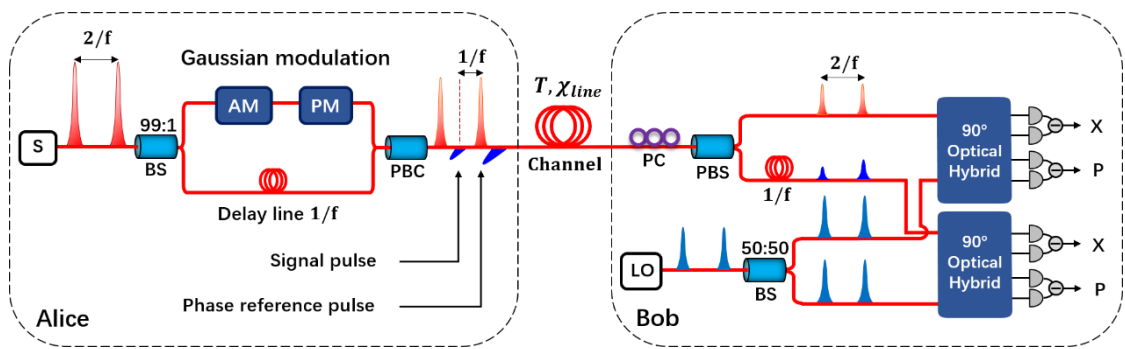

Fig.1. Time-polarization multiplexing LLO CV-QKD scheme. Alice produces a train of coherent state pulses with a repetition of $f/2$ and subsequently splits them into signal pulses and phase-reference pulses using an unbalanced beam splitter (BS). The signal pulses states are Gaussian-modulated, and the phase-reference pulses are delayed by a time $1/f$. Then, a polarization beam combiner (PBC) is used to recombine the signal pulses and the phase-reference pulses into orthogonal polarization modes, sent through the optical fiber which characterized by the channel transmittance T and the channel excess noise $\chi_{line}$. At the receiver, a polarization controller is used to adjust the polarization directions of these two pulses, and the time-polarization multiplexing pulses are split by a polarization beam splitter (PBS), the delay for the signal pulse is $1/f$. Meanwhile, the LO is locally generated and split into two beams for coherent detection of the signal pulses and the phase-reference pulses.

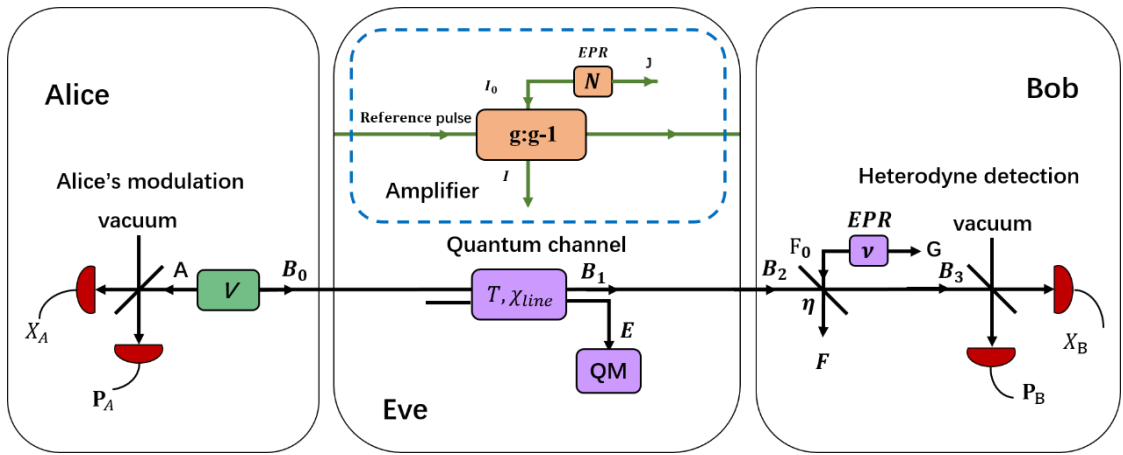

Fig. 2. Phase-reference pulse intensity attack scheme using a phase-insensitive amplifier on a practical time-polarization multiplexing LLO CV-QKD system. Inset with blue-dashed line shows the model for a practical phase-insensitive amplifier. The QM represents Eve's quantum memory.

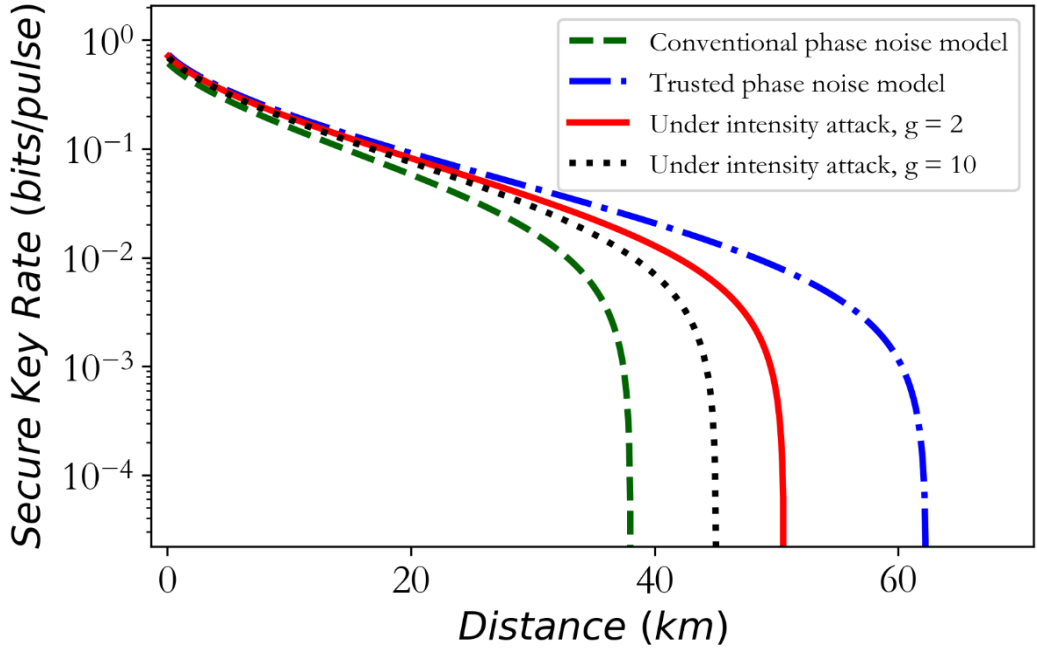

Fig. 3. (Color online). Simulation secret key rate results for the LLO CV-QKD system. The blue dash-dotted line shows the result for the trusted phase noise model, the green dashed line shows the result for the conventional phase noise model. The red solid line and the black dotted line show the results for the trusted phase noise model under the phase-reference pulse intensity attack (intensity attack for short) with the amplification factor of g takes the values 2 and 10, respectively. The noise of the phase-insensitive amplifier $N$ is set to 1 for minimal (vacuum) noise, and the values of the other parameters are as follows: reconciliation efficiency $\beta = 95\%$, detector efficiency $\eta = 0.5$, modulation variance $V_A = 4$, electronic noise $v_{el} = 0.1$, attenuation coefficient $\alpha = 0.2$ dB/km, phase-reference pulse intensity $E_R^2 = 1000$, system excess noise $\xi_0 = 0.01$, ADC quantization number n=10, AM dynamics $d_{dB} = 40$, finite extinction ratios $R_e = 40$ dB and $R_p = 30$ dB. All noise variances here are expressed in shot noise units (SNU).

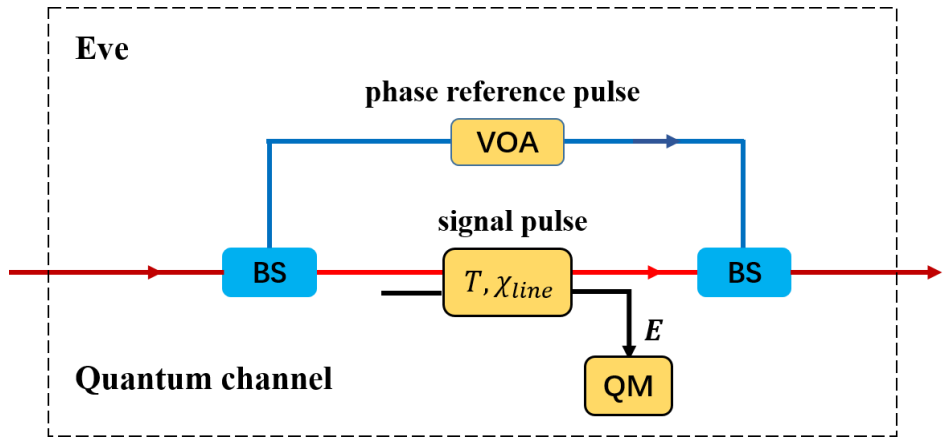

Fig. 4. Phase-reference pulse intensity attack scheme using a variable optical attenuator (VOA). See text for details.

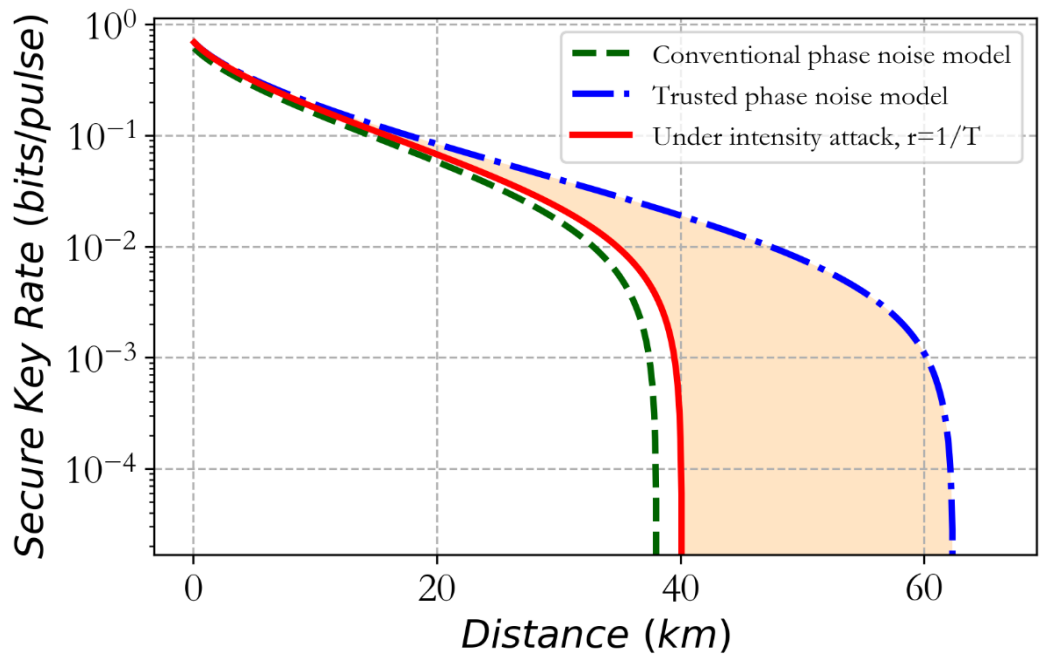

Fig. 5. (Color online) Simulation secret key rate results for the LLO CV-QKD system. The green dashed line shows the result for the conventional phase noise model, the blue dash-dotted line show the result for the trusted phase noise model, and the red solid line shows the result for the trusted phase noise model under the phase-reference pulse intensity attack with $r$ takes the value $1/T$. The bisque region is regarded as the insecure region caused by the attack. The other simulation parameters are the same as that in Fig. 3.

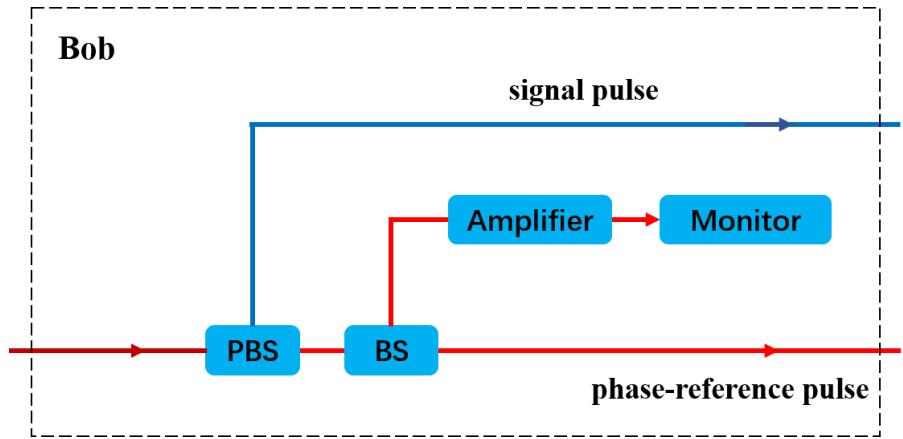

Fig. 6. Monitoring schematic for the phase-reference pulse intensity at the receiver side.